# Numerical simulation of a temporary repository of radioactive material

### E. de la Cruz-Sánchez, J. Klapp*, E. Mayoral-Villa, R. González-Galán, A. M. Gómez-Torres

Instituto Nacional de Investigaciones Nucleares, Carretera México-Toluca S/N, La Marquesa, Ocoyoacac, 52750, Estado de México, México
E-mail: eduardo.delacruz@inin.gob.mx, jaime.klapp@inin.gob.mx, estela.mayoral@inin.gob.mx, armando.gomez@inin.gob.mx, tecnologico59@prodigy.net.mx

*Corresponding autor

### C. E. Alvarado-Rodríguez

División de Ciencias Naturales y Exactas, Universidad de Guanajuato, Noria Alta S/N, Guanajuato, México, and ABACUS-Centro de Matemáticas Aplicadas y Cómputo de Alto Rendimiento, Departamento de Matemáticas, Centro de Investigación y de Estudios Avanzados CINVESTAV-IPN, Carretera México-Toluca Km 38.5, La Marquesa, Ocoyoacac, Estado de México, 52740 México
E-mail: iqcarlosug@gmail.com

**Abstract:** The use of computer simulations techniques is an advantageous tool in order to evaluate and select the most appropriated site for radionuclides confinement. Modelling different scenarios allow to take decisions about which is the most safety place for the final repository. In this work, a bidimensional numerical simulation model for the analysis of dispersion of contaminants trough a saturated porous media using finite element method (FEM), was applied to study the transport of radioisotopes in a temporary nuclear repository localized in the Vadose Zone at Peña Blanca, Mexico. The 2D model used consider the Darcy's law for calculating the velocity field, which is the input data for in a second computation to solve the mass transport equation. Taking into account radionuclides decay the transport of long lived U-series daughters such as $^{238}$U, $^{234}$U, and $^{230}$Th is evaluated. The model was validated using experimental data reported in the literature obtaining good agreement between the numerical results and the available experimental data. The simulations show preferential routes that the contaminant plume follows over time. The radionuclide flow is highly irregular and it is influenced by failures in the area and its interactions in the fluid-solid matrix. The resulting radionuclide concentration distribution is as expected. The most important result of this work is the development of a validated model to describe the migration of radionuclides in saturated porous media with some fractures.

**Biographical notes:**

***E. de la Cruz-Sánchez*** *holds a Masters Degree in Computer Sciences from the Technological Institute of Toluca (1997), with experience in radionuclides migration simulation and in research projects evaluation. He is the manager of the basic science division of the Instituto Nacional de Investigaciones Nucleares (ININ), Mexico.*

***J. Klapp*** *is the Head of the ININ Computational Fluid Dynamics Group and co-founder of the Abacus Centro de Matemáticas Aplicadas y Cómputo de Alto Rendimiento CINVESTAV-IPN Ph. D. in Oxford University, England (1982) and M. Sc. in Cambridge University (1977). He has worked in Computational Fluid Dynamics applied to several fields that range from astrophysics to multiphase flow, dispersions of contaminants in water, soil and the atmosphere, tsunamis, brain simulations and other fields.*

***E. Mayoral-Villa*** *is a Researcher at the "Instituto Nacional de Investigaciones Nucleares" (ININ). She is has a PhD in Sciences (2005), and has a MSc in Physical Chemistry (2001) and graduated in Chemistry (1998) at Universidad Nacional Autónoma de México (UNAM). She has worked in the areas of dynamic systems, renormalization group, colloidal stabilization, poly electrolytes and soft matter, complex fluids, dispersion of radioactive waste and design of materials for nuclear applications. His expertise area is in multiscale numerical simulation using Dissipative Particle Dynamics, Smoothed Particle Hydrodynamics, Molecular Dynamics and* ab initio *DFT.*

***R. González-Galán*** *is the subdirector of support and academic development of the Tecnológico de Estudios Superiores de Tianguistenco, México.*

**Armando Gómez-Torres** earned his Doctor's degree from the Technical University of Munich in 2011. Additionally, he has a BSc in Physics and Mathematics and a MSc in Nuclear Engineering from the National Polytechnic Institute in Mexico. Currently, he works as a Researcher in the National Institute for Nuclear Research in Mexico and is part-time Postgraduate Professor at the National Polytechnic Institute.

***C. E. Alvarado-Rodríguez*** *studied a Ph. D. at the Universidad de Guanajuato,*

*México, and is currently a Postdoc at the Instituto Nacional de Investigaciones Nucleares, México.*

1   **Introduction.**

The operation of nuclear reactors produces burned nuclear fuel, which when removed from the reactor core, needs to be securely managed. The spent fuel could be considered as a waste under some conditions, but in other cases is a prospect for energy production. In both cases, management options (for direct disposal or reprocessing) involve a number of steps, which will inevitably include storage of the burned fuel for a time. This could be short (only few months) or long (around of many decades) depending on the kind of radionuclide. Additionally the natural exploitation of uranium mines produces nuclear wastes that must be confined in an adequate way.

The confinement of radioactive wastes (radionuclides) in underground installations has to take into account its influence on the environment, the solvent action and the waste groundwater drag. One of the main concerns on the lookout for creating a geological repository from high-level nuclear waste (HLW) is the risk that hazardous radionuclides could break away from engineered barriers and be transported to the nearby surroundings. Movement along fractures in the host rock of a repository is one of the main mechanisms by which unsafe radionuclides can be transported to the nearby environment.

The burned fuel for a nuclear reactor is mainly composed of $UO_2$, the remainder corresponds to fission products and actinides produced at some stage of the reactor operation. The solubility of $UO_2$ is particularly small in reducing circumstances. Nevertheless, the contact with oxidizing solutions affects the stability of $UO_2$, producing U(VI) species with superior solubility (Shoesmith, Sunder, Bailey and Wallace, 1989). Soluble species can migrate trough the soil and affect the surroundings.

Computer simulation tools are useful in the analysis of confinement sites for nuclear products. This prevents the unnecessary exposition to radioactive material and gives an estimation of what would happen if an accident occurs. It is important to point out that, especially in the radioactive waste management field, it is essential to discriminate between realistic (quantitative) and conservative (qualitative) models. A realistic understanding, utilizing natural analogues, should be developed and used for improving the calibration of the models with the experimental data available. The incorrect assumptions and simplifications in the model based on experimental data could result in individual conclusions (for analogue studies) which are erroneous and potentially unsafe if used directly in safety assessments. By example, a complex models have been reported in the literature considering the use of naturally occurring U-Th series dispersion in rock-water systems for evaluating the *in situ* retardation of radionuclide migration (Ku, Luo, Goldstein et al., 2009), but discrepancies in these kinds of models and the possibility to use them in general systems have been pointed out (McKinley and Alexander, 1996).

For these reasons, in this work, a numerical study following a traditional model (Bear, 1979) is proposed and, by means of available experimental data, validated. The validated model is then applied to a particular (hypothetical scenario) study case where the same characteristics in the site are present. For this we consider the information of site in the state of Chihuahua, Mexico, where from 1970 to 1980 operated an uraniferous mine which produced about 35 000 tons of uranium tailings (Rojas Martínez, 1996). The waste and naturally occurring minerals, motivated a series of studies with the aim of evaluating the mobility of radionuclides and their possible impact to the surrounding areas. The evaluation concerning to the effectiveness and safety of this repository is an important step in the study of a future permanent disposal in this area. In recent decades there has been a growing need to perform this type of study to define appropriate strategies including adequate barriers for radionuclides disposal. Several works (Neretnieks, 1980; Pollock, 1986; Walton, 1994) focused its attention on the subsurface transport of radionuclide and the hydrodynamic behavior of the barriers on confinement sites.

## 2 Characterization of the site under study.

In this work we consider a temporary repository constructed in the north of Mexico (Chihuahua) in a region known as Peña Blanca (Pearcy, Prikryl and Leslie, 1995; Peterman and Cloke, 2002; Pearcy, Prikryl, Murphhy and Leslie, 1994). This deposit is located in a basin and range horst composed of welded silicic tuff. Uranium mineralization is present in a chemically oxidizing and hydrologically unsaturated zone of structural block. Primary uraninite ($UO_{2+x}$) has altered almost completely to a suite of secondary uranyl minerals (Dobson, Fayek, Goodell et al., 2008; Goldstein, Abdel-Fattah, Murrell et al., 2010). Many experimental studies have been done around this zone, especially in the Nopal I uranium mine, because its geophysical characteristics are very similar to the proposed U.S. high-level nuclear waste (HLW) repository at Yucca Mountain, Nevada (Pearcy, Prikryl and Leslie, 1995; Peterman and Cloke, 2002; Pearcy, Prikryl, Murphhy and Leslie, 1994; Dobson, Fayek, Goodell et al., 2008; Goldstein, Abdel-Fattah, Murrell et al., 2010).

Petrographic analyses indicate that the residual Nopal I uraninite is fined grained (5-10 [mm]) and has a low trace element content (average about 3 wt %). These characteristics compare well with spent nuclear fuel. For an extensive revision of the characterization and experimental studies of this site see Pearcy, Prikryl and Leslie (1995), Peterman and Cloke (2002), Pearcy, Prikryl, Murphhy and Leslie (1994), Dobson, Fayek, Goodell et al. (2008), Goldstein, Abdel-Fattah, Murrell et al. (2010), Goodell (1981), and Leslie, Smart and Pearcy (2005).

Additionally, near this site, a temporary repository has been constructed in order to contain radioactive residues (uranium tiles) originated by the mine exploitation in this area. Leaching of U mill tailings can cause the contamination of soils and aquifers, with the development of groundwater plumes with high concentrations of dissolved U and other metal ions. Risk assessments must be conducted for many of these contaminated sites to evaluate remediation and cleanup scenarios, and a significant component of such risk assessments includes predictions of U transport to drinking water supplies or biological receptors via a groundwater pathway.

In the present work, transport calculations of radionuclides in this temporary repository are performed. For this purpose, the equations that describe the water flow that is seeping through the ground are numerically solved. A classical approach for the solution of this problem (Bear, 1979) consists of considering the porous medium as a continuum and the point-to-point spatial variations of hydrogeological characteristics as an average over a representative element of the volume (REV). The applicability of this approach has been studied by some authors (Long, Remer, Wilson and Witherspoon, 1982; Schwartz, Martys, Bentz et al., 1993). Water hydrodynamic is calculated from a conservation equation based on Darcy´s law. On the other hand, the transport equation considers that the variation of the radionuclides concentration is due to advection, molecular diffusion and mechanical dispersion. This equation also takes into account that the radionuclide which is dissolved in water or adsorbed by the solid matrix may decay. This work shows the mathematical formulation of the model for the simulation of the transport of radionuclides through the ground and presents results for a specific site. The model is validated with experimental data and then it is applied in the analysis of different scenarios in a temporary repository.

**3 Numerical Model.**

For the development of the model some assumptions were stablished:

1. The soil is considered as a fixed continuum porous matrix and the fluid as a mobile continuum phase.

2. The radioactive contaminant (solute) is dissolved in the liquid phase (considering its capacity to dissolve itself in water) constituting collectively a continuum phase, which travel across the solid matrix.

3. The fixed continuum serves as the structure through which the mobile phase flows, and considers the retention of the solute in the media.

4. It is assumed that nuclides, in the dissolved pool, exchange reversibly and quickly with those in the sorbed pool residing on surfaces and with those in the colloidal pool. The exchange is rapid and complete such that a constant ratio of radionuclide concentration exists between the dissolved and sorbed/colloidal pools – as defined by the distribution coefficient, $k_d$.

Taking into account these considerations, the numerical model which is described next was solved. Considering a saturated media, and using Darcy´s law, the dynamics of the fluid (governed by the continuity equation) can be written in the form

$$\frac{\partial}{\partial t}(\rho_f \theta_s) + \nabla \cdot \rho_f \left[ \frac{k}{\eta} (\nabla p + \rho_f g \nabla y) \right] = \rho_f Q_s, \quad (1)$$

where $\theta_s$ is the fraction of the soil occupied by the fluid, $\rho_f$ [$kg/m^3$] is the density of the fluid with the solute, $y$ [$m$] is the vertical coordinate and $p$ [$Pa$] the pressure. The permeability of the porous medium is $k$ [$m^2$], $\eta$ [$Pa \cdot s$] is the viscosity of the fluid, and $g$ [$m/s^2$] the gravity acceleration. $Q_s$ represents the sinks and/or sources of the fluid in the simulation domain.

The term $\boldsymbol{u} = \nabla p + \rho_f g \nabla y$ (equation 1) is called "Darcy´s velocity" and corresponds with the advective term in the transport equation.

With these considerations, the leading equation for a saturated porous medium with water, taking into account the solute retention on the solid matrix and radioactive decay is

$$\theta_s \frac{\partial C_i}{\partial t} + \rho_b \frac{\partial C_{pi}}{\partial C_i} \frac{\partial C_i}{\partial t} + \nabla \cdot \left[ \theta_s \boldsymbol{D}_L \nabla C_i + \boldsymbol{u} C_i \right] - \theta_s \frac{\ln 2}{\lambda_i} C_i - \rho_b \frac{\ln 2}{\lambda_{pi}} \left( \frac{\partial C_{pi}}{\partial C_i} \right) C_i = f, \quad (2)$$

The liquid radionuclide concentration (expressed in mass per liquid volume) of the *i*-th species present in the system is represented by $\theta_s C_i$, and $C_{pi}$ is the adsorbed radionuclide concentration per unit mass in the solid matrix. The hydrodynamic dispersion tensor is given by $\boldsymbol{D}_L$, and $\rho_b$ is the "bulk" density of the porous medium. The $\boldsymbol{u}$ velocity of the fluid is given by equation (1). The radioactive decay constant in the liquid phase and in the solid phase are given by $\theta_s$ (ln $2/\lambda_i$) and $\rho_b$(ln $2/\lambda_i$ )($\partial C_{pi}/\partial C_i$), respectively. The term *f* is the solute source.

Equation (2) establishes that the time variation of the concentration of the radionuclide is mainly due to the following processes:

1. The advection $j_A = \boldsymbol{u}\ C_i$ through which the dissolved radionuclide are dragged by the moving groundwater.

2. The molecular diffusion $\boldsymbol{j}_d = -\theta_s\, D_d\, \nabla C_i$ due to the concentration gradients and determined by Fick´s law, where $D_d$ is the molecular dispersion coefficient.

3. The mechanical dispersion, $\boldsymbol{j}_m = -\theta_s\, \boldsymbol{D}_m\, \nabla C_i$, which acts diluting and reducing the concentration of the solute dissolved in groundwater.

The components of the mechanical dispersion tensor, $\boldsymbol{D}_m$, are defined by the velocity of the fluid and two types of scattering: the longitudinal dispersion that is the scattering that occurs along the fluid streamlines, and the transversal dispersion that it is the dilution that occurs normal to the pathway of the fluid flow. In equation 2, the diffusion and the mechanical dispersion are combined in a single term composed by the mechanical dispersion tensor and the diffusion coefficient to form the hydrodynamics tensor $\theta_s\, \boldsymbol{D}_L = D_d\, \boldsymbol{I} + \theta_s \boldsymbol{D}_m$.

On the other hand, the interaction of the radionuclide with the porous matrix is established by the retention of dissolved radionuclide by the granules that make up the solid matrix. The proportion of the retained to dissolved radionuclide concentration is determined by the relationship between phases: $k_d\, C_{pi} = C_i$, where $k_d$ is the distribution coefficient that describes how much the radionuclide is retained by the solid phase, and how much remains in the liquid phase. The terms $-\theta_s\, (\ln 2/\lambda_i)$ and $-\rho_b\, (\ln 2/\lambda_i)\, (\partial C_{pi}/\partial C_i)$ corresponds to the radioactive decay of each species in the fluid phase, and in the solid phase (absorbed), respectively. $\lambda_i$ is the radioactive half-life constant of the *i-th* specie.

In this work, the radioactive decay chain is studied. That means the disintegration of a radionuclide, so-called "father", which produces elements (daughters) that are also unstable and will disintegrate. The reaction chain is written as:

$$^{238}\text{U} \rightarrow {}^{234}\text{U} \rightarrow {}^{230}\text{Th} \rightarrow {}^{226}\text{Ra} \rightarrow {}^{206}\text{Pb}. \qquad (3)$$

In reality, the decay chain does not directly transmute as it is indicated from $^{238}$U to the stable element $^{206}$Pb, (there are several intermediate decay steps) however, in this work a secular equilibrium process is considered, because the decay time of $^{238}$U, $^{234}$U and $^{230}$Th half-life are much longer than the decay time of the $^{226}$Ra intermediate decay products, and due to that, they are neglected.

## 4  Model Validation.

In order to validate the accuracy of the proposed model, theoretical and experimental results reported in the literature were used as is briefly described hereafter.

### *4.1 Study Case 1: The Nopal I Uranium Deposit in Sierra Peña Blanca, Mexico.*

*4.1.1 Site description.*

As it has been explained, a convenient natural system for estimating the effectiveness of numerical models applying in the study of radionuclide transport is the Nopal I well, localized in the Mexican Peña Blanca (PB-I) uranium district (Peterman and Cloke, 2002; Pearcy, Prikryl, Murphhy and Leslie, 1994; Dobson, Fayek, Goodell et al., 2008; Goldstein, Abdel-Fattah, Murrell et al., 2010; Goodell, 1981; Leslie, Smart and Pearcy, 2005).

In 2003 the PB-I well was drilled at the Nopal I uranium deposit. This well is 250 [m] depth and intersects the local aquifer system penetrating through the tertiary volcanic fragment down to cretaceous limestone. This well has been used for monitoring the increase of the local water table and for sampling groundwater. The PB-I core presents four major stratigraphic units: the Nopal Formation, the Coloradas Formation, the Pozos Formation, and the primary cretaceous limestone basement (see Figure 1). The stratus is briefly described in Table 1 showing the main information for this characterization.

**Table 1** Soil property values (Rojas Martínez, 1996; Green, Rice and Meyer-James, 1995).

| Formation | Lithology | Porosity % | Bulk density [g/cm$^3$] | Permeability [m/s] | Longitudinal diffusion [m] | Transversal diffusion [mm] | $k_d$ [m$^2$/Kg] |
|---|---|---|---|---|---|---|---|
| Nopal | ash-flow tuff | 17 | 2.22 | 6.56304x10$^{-17}$ | 0.05 | 0.001 | $6.96 \times 10^{-2}$ |
| Coloradas | ash-flow tuff | 28 | 2.06 | 1.25494x10$^{-15}$ | 0.05 | 0.001 | $6.96 \times 10^{-2}$ |
| Pozos | Conglo-merate | 15 | 2.22 | 1.26679x10$^{-15}$ | 0.05 | 0.001 | $1.06 \times 10^{-2}$ |
| Limestone | Limestone | 1 | 2.65 | 3.94769x10$^{-18}$ | 0.05 | 0.001 | $3.83 \times 10^{-2}$ |

**Figure 1.** **(a)** Cross section (NE-SW) through the Nopal I deposit, showing the stratigraphic and hydrologic structure for flow and transport models (this figure has been taken for Dobson, Fayek, Goodell *et al.*, 2008. **(b)** Graphical representation for the system. **(c)** Discretization for this system in our model.

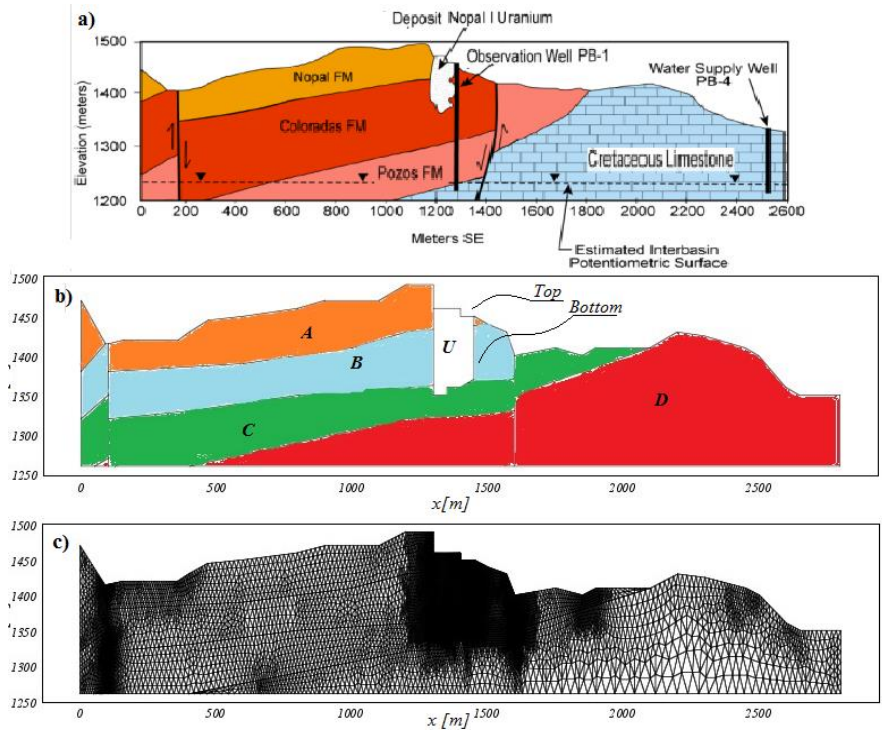

The top component of the PB-1 well is a compactly welded, crystal-rich, rhyolitic ash-flow tuff (Nopal Formation). Breccia zones within the tuff include fractures which have been naturally filled of, limonite, opal, goethite, jarosite and hematite. Following the basic Nopal Formation, the Coloradas Formation is found. This structure consists of a welded lithic-rich rhyolitic ash-flow tuff. Next to this, the underlying Pozos Formation is presented at a depth of 136.38 [m]. This structure contains clasts of subangular to subrounded remains of volcanic rocks, limestone, and chert. At a depth of 244.40 [m] the Pozos formation gets in touch with basic cretaceous limestone basement. The water table is situated at a depth of ~223 [m].

The coefficient of molecular diffusion in soils is always smaller than the coefficient in pure water due the tortuosity. If the coefficient in pure water is of order of $10^{-9}$ m$^2$/s, then the effective molecular diffusion is of order of $10^{-11}$ [m$^2$/s] to $10^{-9}$ [m$^2$/s] (Ingebritsen and Sanford, 1998).

*4.1.2 Model implementation*

The boundary conditions for this model are the following:

1. Pluvial water precipitation in the region enters the model through the upper boundary, this is represented by the Neumann condition

$$\boldsymbol{n} \cdot [\nabla p + \rho_f g \nabla y] = R$$

    with $R = 344$ [mm/year].

2. Vertical walls are assigned by Neumann condition

$$\boldsymbol{n} \cdot [\nabla p + \rho_f g \nabla y] = 0,$$

that specifies wall impervious.

3. In the base of the model (to include the effect of a fracture), it is imposed an output flux

$$\boldsymbol{n} \cdot [\nabla p + \rho_f g \nabla y] = N_0,$$

with $N_0 = 0.01\, K_s$. This value represents a small leak in the base of the model and corresponds to a fraction of the saturated conductivity $K_s$ of the adjacent subdomains.

For equation 2, it is assumed the following boundary conditions:

1. Upper boundary corresponds to the Neumann condition

$$-\boldsymbol{n} \cdot (\theta_s D_L \nabla C + \boldsymbol{q} C) = 0,$$

that is, there is no radionuclide flux across the surface; at both sides a Dirichlet constraint on the concentration (C=0) was applied.

2. For the base at the lower boundary, zero flux advective condition

$$-\boldsymbol{n} \cdot (\theta_s D_L \nabla C) = 0,$$

is imposed.

3. Initial conditions assume that most of the radionuclide is confined in an area which plays the role of an uranium deposit. In other subdomains the initial uranium concentration is zero.

The partial differential equations that represent the system were solved with the COMSOL Multiphysics software, version 3.5a, which is based on the finite element method.

### *4.1.3. Numerical results for Case 1*

A fine grid is used in regions where the complexity of the model demands it. The final mesh contained 10 539 elements. There are no constraints on the direction or magnitude of the radionuclide flux around the uranium deposit. Continuity boundary conditions in the perimeter around the mine were considered.

Calculation of the flow velocity vector field was first obtained using Darcy's law (equation 1). Then, this solution was used for solving the contaminant transport equation (equation 2). The pressure and velocity vector field during the evolution are shown in Figure 2.

**Figure 2.** Pressure field is indicated with color bar and velocity with streamline.

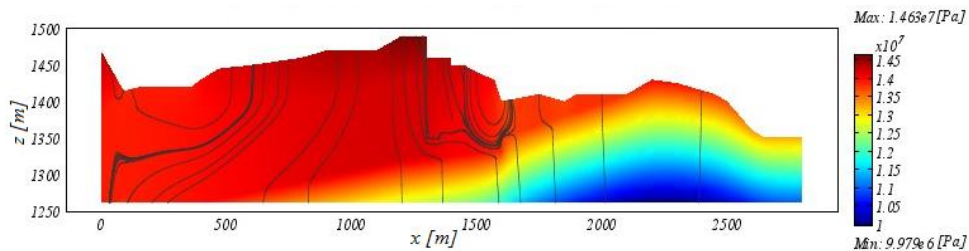

The flow pattern appreciated in figure 2 shows irregular vectors that change with time. The rain water that enters the system, seeps to deeper layers and is influenced by the hydrological properties of the soil that crosses. As is expected, water flow is higher in regions with higher permeability coefficient $k$ (see table 1), and in regions with high pressure gradients.

In the area of the uranium deposit, two main flux patterns can be observed: the first one is located in the top right side of the neighborhood of the mine, and the second in the bottom right side around the mine (see figure 1 b).

The remaining evolution of the dissolved $^{238}U$ contaminant is shown in Figure 3. As mentioned in previous sections, the radionuclide ($^{238}U$) presents in the natural mine, is initially contained in a "cavity" which is indicated in Figure 1 with the letter *U*. The rainwater dissolves the radionuclides present in the soil, and thereafter these radionuclides are dragged to deeper layers following the patterns described by the fluxes shown in Figure 2.

In the model, the initial $^{238}U$ concentration in the deposit is $C_0$=1.475 X $10^{-12}$ [Kg/m$^3$], which is the maximum possible concentration taking into account the solubility for this species (Shoesmith, Sunder, Bailey and Wallace, 1989). After a time $t$ ($^{234}U$) = 4.468 X $10^9$ [years], and $t$ ($^{308}Th$)= 75 380 years, new radioisotopes creation is considered ($^{234}U$ and $^{230}Th$). The time evolution of the daughters $^{238}U$, $^{234}U$ and $^{230}$Th are also calculated. The change in the concentration of the radionuclides by advection, diffusion, and radioactive decay over the course of three millions years were simulated. After an initial period of zero contaminant, the evolution in the process distributes the radionuclide in the different substrates according with its geological characteristics.

**Figure 3**. Evolution of the pollutant concentration $^{238}U$ [kg/m$^2$] with time [s]. In panel (a): full system, bottom panels: detail of the figure around the uranium deposit to times: (b) plume of $^{238}U$ at $t = 2 \times 10^6$ [years]; (c) $t = 2.5 \times 10^6$ [years] and (d) $t = 3 \times 10^6$ [years].

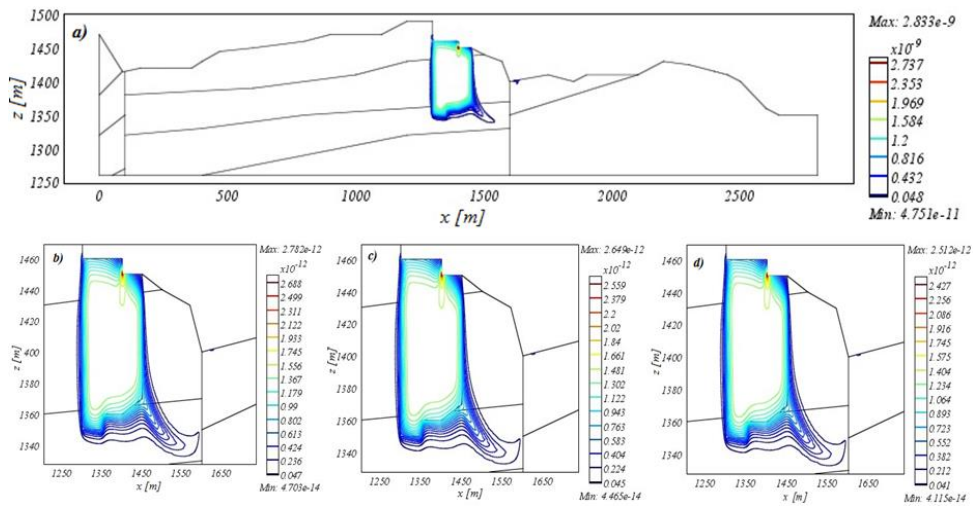

The evolution of the radionuclide concentrations ($^{338}U$, $^{234}U$ and $^{230}Th$) with time at different depths was estimated along five fixed vertical lines at $x_1 = 1\,200$ [m], $x_2 = 1300$ [m], $x_3 = 1400$ [m], $x_4 = 1500$ [m] and $x_5 = 1600$ [m], as shown in figure 4; all plots are for $t = 9.501 \times 10^{13}$ [s]. The evolution shows that the concentrations of the radioisotope increases reaching two maximums, the first one around the vertical coordinate $z = 1\,445$ to $1\,465$ [m], and the second one around $z = 1\,320$ to $1\,370$ [m]. These two maximums corresponds with a depth of 35 [m] and 180 [m] from the surface ($z = 1\,500\,m$), and are in good agreement with Dobson, Fayek, Goodell et al., 2008, where the maximum values (> 50,000 [cps]) are encountered in the upper 15 [m] of the borehole, and correspond to a zone of strong variation and also in the neighborhood of the water table, located at a depth of ~222.6 [m], where there is a slim region (216.6–225.1 [m]) with particularly elevated natural gamma values. These measures are all above 2 500 [cps], with a maximum value of 7 927 [cps]. Under this sector, gamma values go down quickly, and at depths larger than 227.1 [m]*,* all log gamma values are lower than 400 [cps]. This also corresponds with the behavior observed in these simulations because the radioactive plume follows a specific pattern and does not arrive in the region up to this zone.

**Figure 4**. Distribution of radionuclide concentration with depth along of five vertical lines located in $x_1 = 1\,200$ [m], $x_2 = 1\,300$ [m], $x_3 = 1\,400$ [m], $x_4 = 1\,500$ [m], $x_5 = 1\,600$ [m]. All graphics are for $t = 9.501 \times 10^6$ [years]. In panel (a): Radionuclide $^{238}U$; (b) $^{234}U$ and (c) $^{234}Th$.

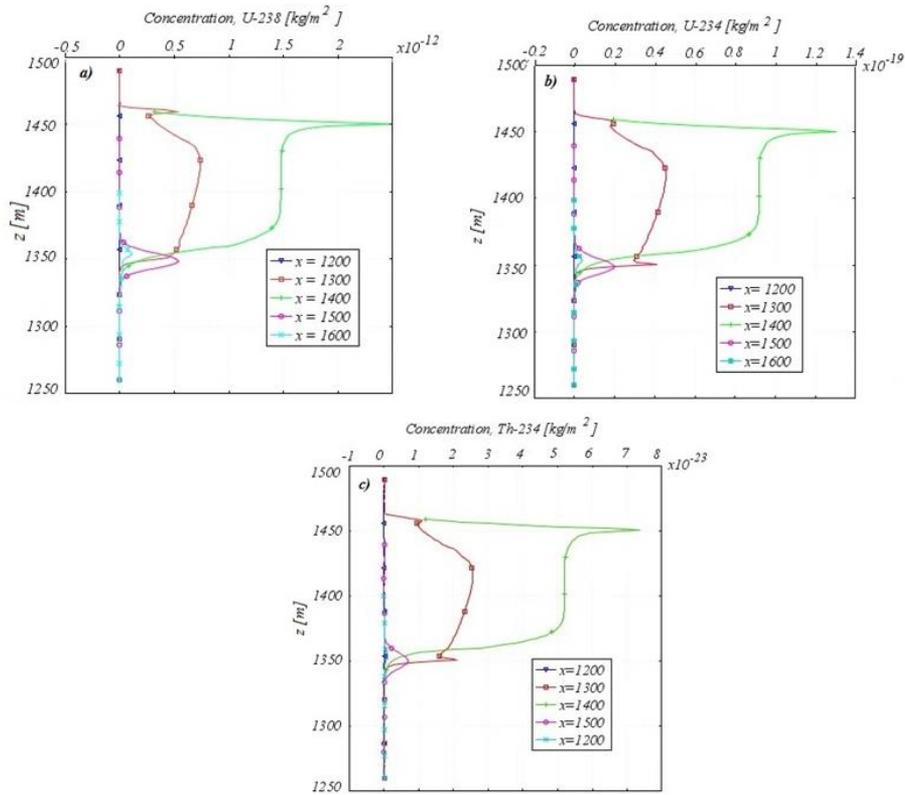

In Figure 5 shows the father/daughter concentration ratio with time. For each item, three lines are plotted which corresponds with the three points of interest $z = 1\,350$ [m], $z = 1\,400$ [m], and $z = 1\,500$ [m]. Similarly, each subsection shows these relations for the horizontal coordinates $x = 1\,450$ [m], $x = 1\,475$ [m], and $x = 1\,500$ [m].

**Figure 5.** Concentrations ratios between parent/daughter: $^{238}U/~^{234}U$ and $^{234}U/~^{230}Th$ over time obtained in the simulation.

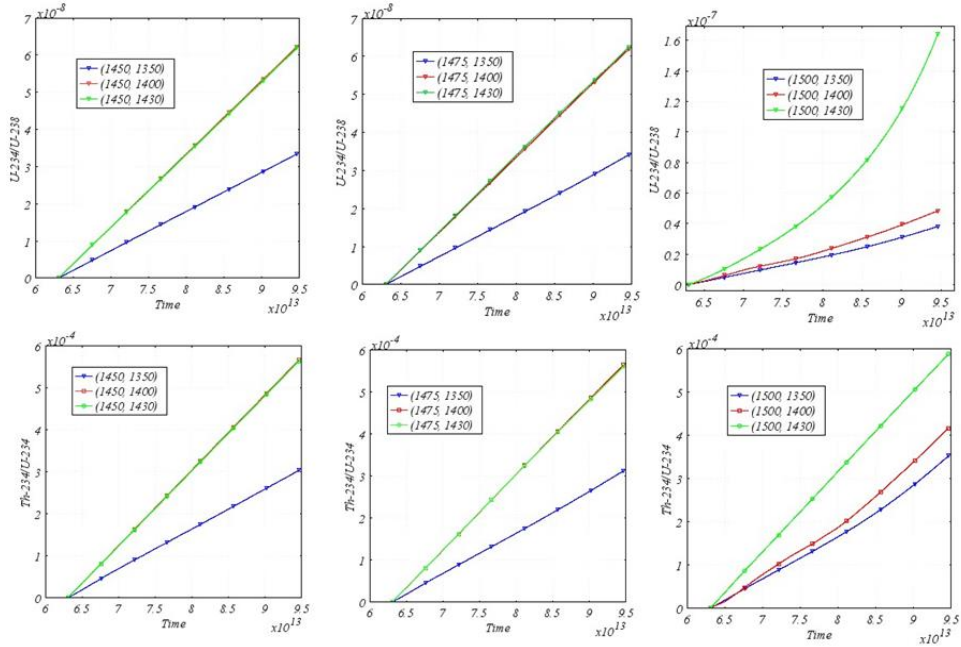

A linear dependence of this quantity is observed and corresponds with the expected behavior observed in laboratory experiments and reported in the literature (Ivanovich and Harmon, 1992).

### *4.2 Case 2. Uranium repository tails in Sierra Peña Blanca, Mexico.*

### *4.2.1. Site description*

In this case, the model has been applied to analyse the radionuclide concentration in the subsurface of the zone where the uranium repository tails in Sierra Blanca is located. Arrival times at the points of interest, shape and direction of the plume were calculated. The site under study has a desert climate with an annual rainfall of 300 [mm] and an average temperature of 18º [C] (Ruiz Cristobal, 1998). The Figure 6 shows a cross-section model of the subsoil layers. In our case six layers are considered identified as *k_C01, k_C02, k_C03, k_C04, k_C05, k_C06*. Table 2 summarizes the hydrogeological parameters of the sub-soils, where *K* represents the hydraulic conductivity or permeability for each strata. $\theta$ is the porosity, and finally $k_d$ is the distribution coefficient.

**Figure 6.** The subsurface strata in the computational domain. The horizontal (*x*) and vertical (*y*) axis are given in meters.

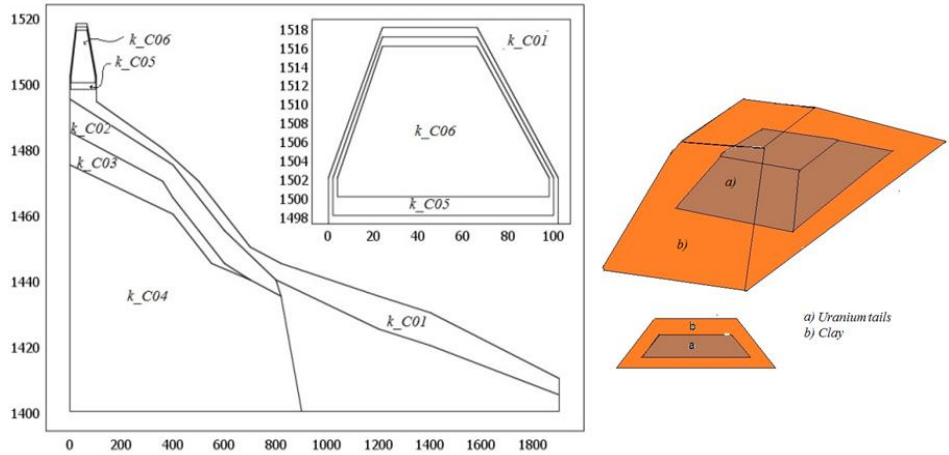

Layer *k_C04* is the most superficial and is composed of a stuff so-called "caliche". The *k_C02* layer is composed of conglomeratic sand, and below both we find the conglomerate layer *k_C03*. The deepest layer, the *k_C04* is composed of volcanic stone and limestone on a sandy matrix. The radioactive wastes were stacked on a clay bed of 1[m] thickness forming a truncated pyramid. The pyramid is covered with a layer of the same clay used in the bed. The whole arrangement is covered with soil of the region.

In this work a fragment of the uranium decay chain was considered. The first specie analyzed is the radionuclide $^{230}$Th, which has a constant half-life of 75 380 [years]. The decay product of thorium is $^{226}$Ra, which in turn decays into a series of elements that also are unstable and have a very short half-life (compared with the half-lives of thorium and radium), and whose final element is $^{206}$Pb which is stable. The intermediate elements between $^{226}$Ra and $^{206}$Pb are not considered, since they are in secular equilibrium.

**Table 2.** Hydrologic parameters of the computational model.

| Stratum | Composition | $K\ (cm^2)$ | $\theta$ | $k_d\ (m^2/Kg)$ |
|---|---|---|---|---|
| *k_C01* | "Caliche" | $1 \times 10^{-5}$ | $\theta_1 = 0.249$ | $6.96 \times 10^{-2}$ |
| *k_C02* | conglomeratic sand | $1 \times 10^{-4}$ | $\theta_2 = 0.2$ | $1.06 \times 10^{-2}\ (k_{d\_salt})$ |
| *k_C03* | Conglomerate | $1 \times 10^{-3}$ | $\theta_3 = 0.3$ | $1.06 \times 10^{-2}\ (k_{d\_salt})$ |
| *k_C04* | "Chontes" | $1 \times 10^{-5}$ | $\theta_4 = 0.05$ | $6.96 \times 10^{-2}$ |
| *k_C05* | Clay | $7.4 \times 10^{-8}$ | $\theta_5 = 0.05$ | $3.83 \times 10^{-2}\ (k_{d\_clay1})$ |
| *k_C06* | Compacted tailing | $7.2 \times 10^{-6}$. | $\theta_6 = 0.01$ | $3.06 \times 10^{-2}\ (k_{d\_clay2})$ |

*4.2.2. Model implementation*

The initial and boundary conditions for the equation 1 are:

1. For the equation 1, the upper boundaries represent the soil surface, so is defined a Neumann condition that describes the inflow of water to the system.
2. The bottom borders ($y = 1\ 400$ [m]) and the left side ($x = 0$ [m]), are impermeable, so is imposed a zero flow Neumann condition.
3. The border at the right side ($x = 1\ 900$ [m]) is a Dirichlet boundary which corresponds with a zero pressure for the area of the exit flow.
4. The boundary conditions for the equation 2 are imposed to indicate that there is no radionuclide flow entering through external borders; this is described by a zero flow Neumann condition.
5. The right side of the model is the output of the system, so is established an advective flow condition because this is the dominant component in flow.

In our model, water enters through the upper boundaries and seeps to the deepest layers. When the water arrives to the *k_C06* stratum, the radionuclide is dissolved and dragged to deeper strata which initially are non-contaminated. The following considerations and parameters were used:

1. Through the external boundaries of the model there is no radionuclide inflow, so that the only source of contamination is the subdomain 6, while other subdomains are initially free of radionuclides.
2. It is considered that the movement of radionuclide through subdomain 6 is given only by diffusion.
3. The amount of radionuclide that migrates outside subdomain 6 is a function of the amount of substance that can be dissolving in the water present in the medium.

The process of dissolution of radionuclides can be complex and are governed by many reactions such as: radiolysis, complex formation and chemical dissolution independent of the effects of radiation. In this paper only the latter process was considered by defining the dissolution rate applied to the boundaries of subdomain 6. The time evolution of the concentration at each point depends on the processes of mechanical dispersion, molecular diffusion and advection. The maximum simulation time is 2 million years. In our model we use total concentrations for every chemical species; this allows that the amount of variables does not depend of the number of minerals that locally precipitated.

*4.2.3. Numerical results for Case 2.*

The flow velocity field calculation is done using Darcy's law. Then the solution is used for solving the contaminant transport equation. This is possible only when the concentrations of the solutes are such that does not affect the density and viscosity of the fluid. In our case this is true because the concentrations that have been evaluated are very small. The fluid discharge velocities obtained for each subdomain are contrasted with the velocities reported by Rojas Martínez, 1996, see Table 3. The order of magnitude of the velocities in our model corresponds with the values reported however we also obtain the velocity flow field and the detailed radioactive transport.

**Table 3.** Fluid discharge velocities (Taken from Rojas Martínez, 1996)

| Stratum | Velocity reported [m/s] (Rojas Martínez, 1996) | Velocity max/min calculated [m/s] (our work) |
|---|---|---|
| $k\_CO1$ | $10^{-7}$ | $8.21 \times 10^{-7}$ |
|  | $10^{-9}$ | $8.098 \times 10^{-12}$ |
| $k\_CO2$ | $10^{-6}$ | $3.401 \times 10^{-6}$ |
|  | $10^{-7}$ | $1.625 \times 10^{-9}$ |
| $k\_CO3$ | $10^{-5}$ | $6.315 \times 10^{-6}$ |
|  | $10^{-7}$ | $1.24 \times 10^{-9}$ |
| $k\_CO4$ | ----- | $1.232 \times 10^{-7}$ |
|  |  | $1.149 \times 10^{-12}$ |

The plots of figure 7 shows the distribution of the $^{230}$Th (left panel) and $^{226}$R (right side) concentration *vs* depth for the time = 2 millions of years. The maximum magnitude of the concentration is in the repository zone ($y = 1\,500$ [m]) and then decreases very fast in range between $y = 1\,500$ [m] and $y = 1\,480$ [m]. Moreover, in uranium processing plants we have drainage flows with values below 1 [pCi/m$^3$]. This represents a range of $4 \times 10^{-13}$ to $4 \times 10^{-18}$ [M]. In our case, the maximum concentration of the radionuclide is of the order of tens of pico-grams, and occurs near the area of the repository, which is consistent with the experimental data.

**Figure 7.** Distribution of the contaminant concentration with depths at time = 2 X $10^6$ [years]. In panel (a) concentration of $^{230}$Th. (b) concentration of $^{226}$Ra

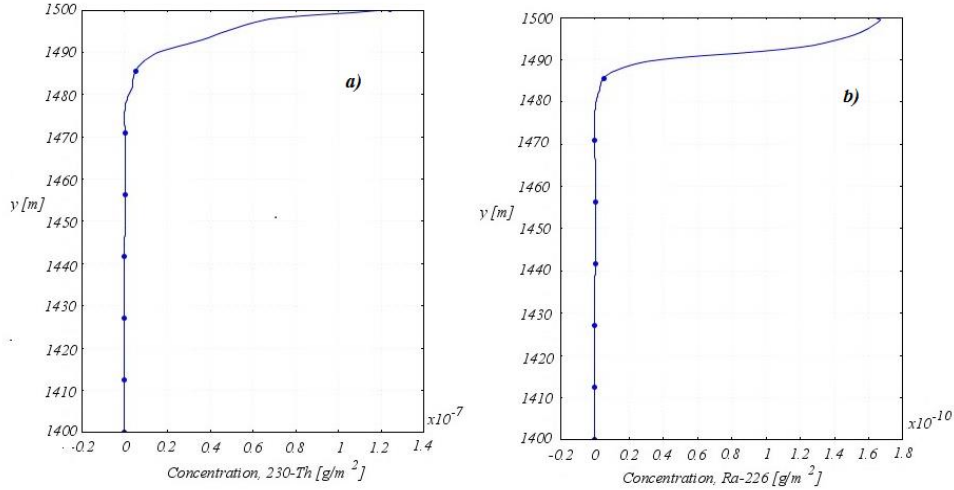

The plots shown in figure 8 are the $^{230}$Th (panels (a) and (b)) and $^{226}$R (panels (c) and (d)) plume for the times: $t = 3.1536$ x $10^{13}$ [s] and (b) $t = 6.3072$ x $10^{13}$ [s]. Our plots show that the contaminant plume has its source in the region near $x =$ 40 [m], $y = 1\,500$ [m], and $x = 100$ [m], $y = 1\,500$ [m], thereafter it moves towards a virtual channel located in the region 60 [m] $< x <$ 85 [m]; the above results suggest a local pressures change, and velocities that conveys the plume to that area.

With the purpose of comparison, we performed the simulation of this system under normal operation (model A) and under a condition of fracture in the engineered barrier (model B).

In this regard, Figure 9 shows the $^{226}$Ra concentration with time for the points: (65, 1 460), (65, 1 470), and (65, 1480). In left side panel, we present model A and in the right side panel, the model with the failure. It is easy to see that in model B the $^{226}$R concentration is higher than in model A.

**Figure 8.** Plume contaminant concentration for the times 1 million years (left panels) and 2 million years (right panels).

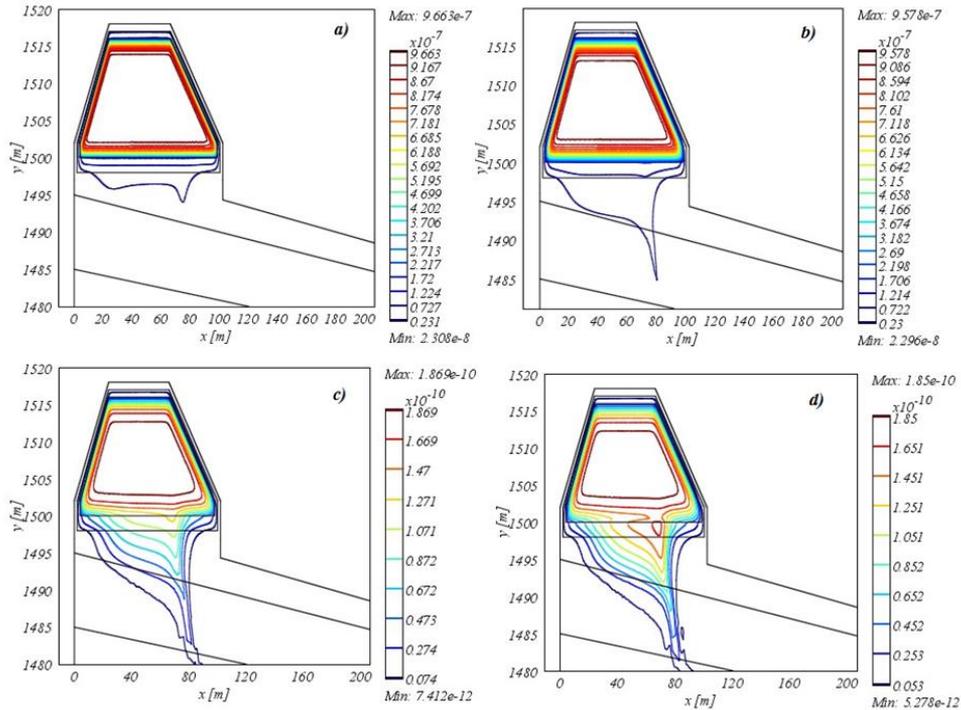

**Figure 9.** a) Concentration of $^{230}$Th with time in model A, b) Concentration of $^{230}$Th along time in model B.

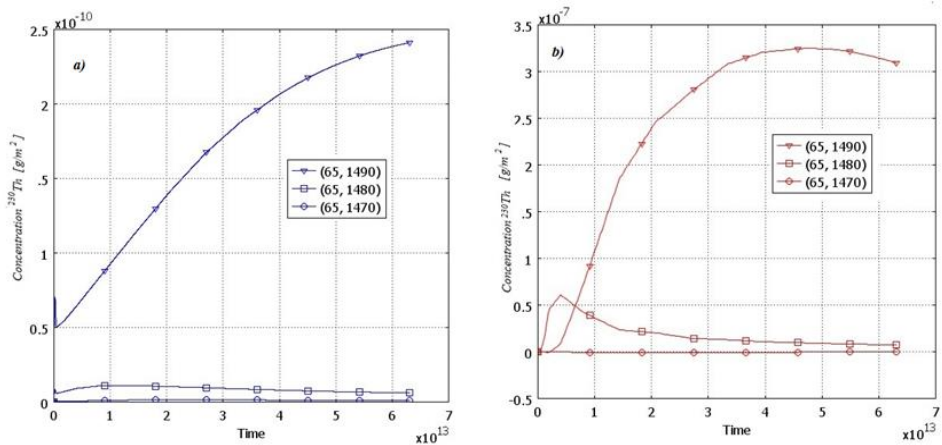

The water that seeps into the subsoil may contain a significant amount of dissolved substances which are called solutes. In the case of underground installations for radionuclide disposal, this is a factor to be taken into account.

## 5      Conclusions

We have developed a numerical model for simulating the diffusion and transport of radioisotopes through a porous media. The numerical model takes into account the phenomena of molecular diffusion, advection, radioactive decay and radionuclide adsorption by the porous matrix.

Our 2D model utilizes the Darcy's law for calculating the velocity field. It is subsequently used as input data for solving the mass transport equation considering the radionuclide decay. The model was validated using experimental data reported in the literature for the Vadose Zone at Peña Blanca, Mexico obtaining very good agreement between our numerical results and available experimental data. With this validated model and parameters we evaluated how the radionuclide would migrate if an eventual failure occurs in the container located in this particular site in the north of Mexico. The simulations show preferential routes that the contaminant plume follows over time. The radionuclide flow is highly irregular and it is influenced by failures in the area and its interactions in the fluid-solid matrix. The obtained concentration of the radionuclide was as expected, Dobson *et al.*, (2008), reported that there are two areas with a higher radionuclide concentration. These zones are located at 15 [m] depth from the soil and near the water table with 226 [m] depth. The fluid velocities that are estimated in our simulations are in a range between $7.551 \times 10^{-15}$ [m/s] and $6.315 \times 10^{-6}$ [m/s]; these values have the same order of magnitude as those reported by (Rojas Martínez, 1996; Ruiz-Cristóbal, 1998).

In uranium processing plants we have drainage flows with values below 1 [pCi/m$^3$]. This represents a range of $4 \times 10^{-13}$ to $4 \times 10^{-18}$ [M] (IAEA Source Book, 1992). In our case, the maximum concentration of the radionuclide is of the order of tens of picograms, and occurs near the area of the repository, which is consistent.

The estimated radionuclide concentration in areas surrounding the landfill site were of the order of picograms, which can be considered a low level. However, the simulation of a failure in engineered barrier produces higher concentrations and a much higher mobility. This suggests that this nuclear waste disposal method can be efficient and inexpensive; however, it is essential to have a careful design of the facilities and also a permanent monitoring to minimize any eventuality.

We observe a linear dependence of the parent/daughter ratio with time, this corresponds with the expected behavior observed in laboratory experiments and reported in the literature (Ivanovich and Harmon, 1992), which is due to the fact that the mobility of both radionuclide is the same in our model.

**Acknowledgements**

This work was partially supported by ABACUS, CONACyT grant EDOMEX-2011-C01-165873.